\documentclass{nature}

\usepackage{graphicx}

\title{Ultrafast Observation of Critical Nematic Fluctuations and Giant Magnetoelastic Coupling in Iron Pnictides}

\author{A. Patz$^{1,2,\dagger}$, T. Li$^{1,2,\dagger}$, S. Ran$^{1,2}$, R. M. Fernandes$^{3}$, J. Schmalian$^{4}$, S. L. Bud'ko$^{1,2}$, P. C. Canfield$^{1,2}$, I. E. Perakis$^{5}$  and J. Wang$^{1,2}$}

\begin{document}

\maketitle

\begin{affiliations}
\item Department of Physics and Astronomy, Iowa State
University, Ames, Iowa 50011, USA
 \item Ames Laboratory - USDOE, Ames, Iowa 50011, USA
\item School of Physics and Astronomy, University of Minnesota, Minneapolis, MN 55116, USA
 \item Institute for Theory of Condensed Matter Physics and Center for Functional Nanostructutes, Karlsruhe Institute of Technology, Karlsruhe, 76131, Germany
\item Department of Physics, University of Crete,
Heraklion, Crete, 71003, Greece,  
 and
Institute of Electronic Structure \& Laser, Foundation for
Research and Technology-Hellas, Heraklion, Crete, 71110, Greece
\end{affiliations}



\begin{abstract}
A significant anisotropy manifests ubiquitously in normal state properties of many of the iron pnictides,
and has emerged as a cross-cutting challenge 
for understanding complexity in advanced materials, quantum magnetism and high-T$_c$ superconductivity. \cite{7TanatarPRB10,1ChuScience10,3YingPRL11,4YiPNAS11,5NakajimaPNAS11,
6DuszaEPL11,16KasaharaNature12,chu2012,Gegenwart13} 
Although an \emph{electronically} driven tetragonal symmetry-breaking (electronic nematicity) has been invoked, its distinction from 
spin and structural
orders is still hindered as they couple together to break the same symmetry.\cite{9SiPRL08,10FangPRB08,11XuPRB08,13FernandesPRL10,15FernandesPRB12,DevereauxPRB10}  
Here we use femtosecond-resolved polarimetry to reveal critical order parameter flucations   
of the nematic state of unstrained Ba(Fe$_{1-x}$Co$_{x}$)$_2$As$_2$. 
The anisotropic response, which arises from the in-plane anisotropy of the refractive index, displays a characteristic two-step recovery absent in 
the lattice and electron dynamics.
The fast recovery appears only in the magnetically ordered state, whereas the slow one persists in the paramagnetic 
phase with a sharp increase of the relaxation time approaching the structural transition temperature, indicative of the critical divergence of nematic fluctuations. Particularly, our results show a gigantic magnetoelastic coupling that far exceeds electron-phonon coupling, a ratio $\sim$10000 times larger than conventional magnetic metals.
This establishes an independent electronic nematic degree of freedom in iron 
pnictides, and bridges the gap between quantum nematic matter \cite{FradkinARCMP10} and 
technologically relevant functionalities. 
\end{abstract}

In the electronic nematic state of the iron pnictides, the high-temperature tetragonal symmetry ($C_4$) of the system is lowered to an orthorhombic one ($C_2$) at the temperature $T_s$ through the divergence of nematic fluctuations (Fig. 1a). The associated nematic order parameter $\phi$ has therefore Ising ($Z_2$) character (two discrete values) \cite{10FangPRB08,11XuPRB08,13FernandesPRL10} and couples to various other degrees of freedom, such as magnetic fluctuations, lattice orthorhombicity, and orbital order \cite{15FernandesPRB12}, giving rise to anisotropies in several quantities -- electrical and thermal transport, orbital occupation, optical conductivity, and magnetic susceptibility.\cite{7TanatarPRB10,1ChuScience10,4YiPNAS11,5NakajimaPNAS11,6DuszaEPL11,
16KasaharaNature12,chu2012,Gegenwart13}. In most materials, the nematic transition usually takes place very close to a magnetic transition at $T_N \leq T_S$ to a (0, $\pi$)/($\pi$, 0) spin-striped state \cite{KimPRB11}. 
The challenge 
arises on how to underpin the origin of the anisotropy from electronic nematicity and disentangle it from various different, yet cooperative, contributions to the anisotropy.

Ultrafast dynamics of the nematic order parameter $\phi$ in iron pnictides has not been measured thus far, despite the fact that this approach can potentially reveal dynamic and critical fluctuations associated with the establishment of the new phase. The time resolution also disentangles the various coupled order parameters based on their distinct relaxation dynamics subsequent to being suddenly driven out of equilibrium, thus revealing information hidden inside the time-averaged signals obtained from static measurements.   
Prior ultrafast experiments in iron pnictides mostly concentrated on laser-induced melting of the superconducting and SDW gaps, and on charge quasi-particle dynamics, but provided little insights on the nematicity since it is not known how to explicitly resolve the $\phi$ dynamics.
\cite{18MansartPRB10,19ChiaPRL10,20TorchinskyPRL10,Merteljprl2009,21StojchevskaPRB12,Kim2012}
Recently, linear polarization spectroscopy from the far-infrared up to 2 eV revealed a static conductivity anisotropy
along the $a$ and $b$ orthorhombic directions (Fig. 1a),
which paves a way to optically probe the $Z_2$ Ising order parameter.\cite{5NakajimaPNAS11,6DuszaEPL11} However, 
two key issues remain before one can access the unexplored ultrafast regime: first, probing anisotropy in the \emph{ultraviolet (UV)} region, far above the Fermi energy,
is most critical since it minimizes the contamination of ultrafast polarization signals by spurious effects, e.g., spectral weight transfer associated with pump-induced Fermi-surface reconstruction (see, e.g., Ref. \cite{21StojchevskaPRB12} and methods).   
Thus far, optical anisotropy above 2 eV could not be observed in conventional broadband polarization optics \cite{5NakajimaPNAS11}.   
Second, to access the ``spontaneous" order $\phi$, it is critical to study unstrained single crystals. Even a small external strain will smear out the well-defined critical temperature, predicted at T$_s$
\cite{1ChuScience10,7TanatarPRB10}.  

This letter characterizes ultrafast relaxation of the Ising-nematic order parameter $\phi$ 
following non-equilibrium photoexcitation at 1.55eV of unstrained Ba(Fe$_{1-x}$Co$_{x}$)$_2$As$_2$ using laser-based, femtosecond (fs)-resolved polarimetry probe centered at the blue/near-UV region at 3.1 eV (methods) \cite{LiNature13,LiPRL12}. 
A two-step polarization recovery with fast ($\tau^{\mathrm{fast}}$) and slow ($\tau^{\mathrm{slow}}$) relaxation times 
is observed only in the anisotropic response, \emph{absent} in the isotropic response that arises from the transient electron and lattice relaxations. 
By studying both parent (x=0.00, with $T_s=T_N=136$K) and underdoped (x=0.047, with $T_s=66$K and $T_N=48$K) single crystals (methods) \cite{CanfieldARCMP10},
we are able to separate the contribution to the anisotropy 
from long-range magnetic order, $\tau^{\mathrm{fast}}$, present only below $T_N$, to critical divergence of the nematic fluctuations, $\tau^{\mathrm{slow}}$, present upon approaching $T_s$.  
A non-equilibrium model \cite{25BeaurepairePRL95} reproduces the fs dynamics,  
and reveals, particularly, a gigantic ratio between the magnetic/nematic-phonon and electron-phonon couplings, $\sim$10000 times larger than conventional magnetic metals.

We first demonstrate a laser-based polarimetry technique centered at 3.1 eV that directly measures the static nematic order parameter $\phi$.
As illustrated in Fig. 1b , when a linearly-polarized optical field, oriented at 45 degrees with respect to the orthogonal axes of different complex refractive indices $\widetilde{N}_{a}$ and $\widetilde{N}_{b}$, is incident on the sample surface, an elliptically polarized light ($\eta$) with a rotation of the polarization plane ($\theta$) is observed to be reflected (methods). A complex angle $\widetilde{\Theta }=\theta +i\eta$ can be obtained by expanding the solution of the Fresnel equation with respect to $\alpha =\Delta\widetilde{N}/\widetilde{N}$ (defined as ($\widetilde{N}_a-\widetilde{N}_b)/(\widetilde{N}_a+\widetilde{N}_b)\ll$1):
\begin{equation}
\widetilde{\Theta }=\frac{2\Delta\widetilde{N}}{\widetilde{N}^{2}-4}+O(\alpha ^{2}) 
\end{equation}
and is directly proportional to $\Delta\widetilde{N}$, i.e. $\widetilde{\Theta} \propto \phi$.
Since the real and imaginary parts of $\widetilde{\Theta}$ are connected via Kramers-Kronig relations we can then focus only on the imaginary part of $\widetilde{\Theta }$, $\eta$ to extract information about $\phi$.  We obtain $\eta$ by subtracting the $s$ and $p$ polarized components $(I_s-I_p)/2(I_s+I_p)$, and also probe the surface reflection $R$=$I_s+I_p$, a measure of the isotropic response (black box). The result for unstrained Ba(Fe$_{0.953}$Co$_{0.047}$)$_2$As$_2$ (black dots) is shown in Fig. 1c. The $\eta$ shows a gradual increase only below $T_s=66$K and exhibits a distinct increase across $\approx T_N=48$K, revealing the coupling between the nematic order parameter $\phi$ and the magnetic and orthorhombic order parameters obtained with x-ray and neutron scatterings (Fig. 1d).\cite{23NandiPRL10} 
A \emph{quantitative} comparison of the temperature profile of $\eta(T)$ shows a good agreement with the theoretically predicted behavior of $\phi(T)$. 
In this calculation, the degeneracy of the magnetic ground state, allied to magnetic fluctuations, gives rise to a non-zero nematic order parameter at $T_s$, which is enhanced at $T_N$ due to the feedback effect on the magnetic spectrum (see Fig. 1a and methods).
Notice that the full temperature dependence of $\phi(T)$ has been inaccessible in various probes, e.g., due to the inapplicability of magnetometry in the SDW phase \cite{16KasaharaNature12} and due to the limited signal-to-noise ratio in polarized photoemission.\cite{4YiPNAS11} 

The fs-resolved ellipticity change $\Delta\eta$ in the parent compound is shown in a logarithmic scale as a function of time delay in Fig. 2a at 4K.
Ultrafast photoexcitation results in softening of the nematic order, revealed by a negative change $\Delta\eta<$0, which is followed by a bi-exponential recovery composed of an initial fast relaxation of $\tau^{\mathrm{fast}}\sim$1.2ps and a slow one of $\tau^{\mathrm{slow}}\sim$28ps. 
The most salient feature in Fig. 2 is the striking difference in temporal profiles between the polarization $\Delta\eta$ and simultaneously-probed reflectivity $\Delta R/R$. 
This difference persists both in the fs and in the subsequent extended 100 ps time scales. Shown in the inset of Fig. 2a is the comparison for the first 1.2 ps, which reveals a delayed rise $\Delta\tau_{rise}$ of $\Delta\eta$ (black dots) compared to $\Delta R/R$ (red shade),  indicating a faster electron thermalization $\sim$150fs prior to the softening of the nematic order. 
Three key properties of $\tau^{\mathrm{fast}}$ and $\tau^{\mathrm{slow}}$ are noted following the initial thermalization: (i) both of them are absent in the $\Delta R/R$ decay profiles, determined by cooling of hot electrons and lattice since $\Delta R/R$ originates from thermally-induced band structure renormalization. 
This remarkable difference in the time evolution of the $\eta$ strongly points to a separate physical origin of the former: it is independent of either electron or structural dynamics.   
(ii) The $\tau^{\mathrm{fast}}$ and $\tau^{\mathrm{slow}}$ components exhibit distinctly different dependence on temperature (Figs. 2b and 2c): $\tau^{\mathrm{fast}}$ displays little variation between 4K and 130K traces, while $\tau^{\mathrm{slow}}$ shows a much faster relaxation at low temperatures (T=4K) than at temperatures (T=130K) slightly below T$_s$ (=136K).
A detailed fitting of temperature dependence of the two-step recovery (red lines), shown in Fig. 2d, reveals that the $\tau^{\mathrm{slow}}$ is mostly constant up to near T$_s$ where it sharply increases. This behavior indicates a critical divergence near the phase transition. 
In contrast, $\Delta R/R$ mostly exhibits a {\em temperature-independent} decay profile: the 4K and 130K traces (red shade, Figs. 2b-2c) are almost identical. These data rule out any single particle electron or pure structural, i.e., phonon, origin for the critical slowing down in $\tau^{\mathrm{slow}}$, and therefore points to a divergence of nematic correlation at T$_s$ \cite{15FernandesPRB12,chu2012}. 
iii) The photoinduced $\Delta\eta$ amplitude quickly diminishes above T$_S$, but unlike for $\Delta R/R$, again corroborating the nematic origin of the observed transient $\eta$ signal. Note that $\Delta R/R$ also has a periodic oscillation of $\sim$120 fs, which arises from the As-As optical phonon \cite{18MansartPRB10}.  

In order to differentiate the correlation mechanisms leading to the $\tau^{\mathrm{fast}}$ and $\tau^{\mathrm{slow}}$ components, Figs. 3a-3b present ultrafast $\Delta\eta$ measurements on the x=0.047 sample. 
At 4K, we observe very similar bi-exponential relaxation behavior as in the x=0.00 sample. 
However, at T=54K slightly above T$_N$=48K, the $\tau^{\mathrm{fast}}$ component disappears, while a {\em similar} $\tau^{\mathrm{slow}}$ relaxation persists. 
Detailed temperature dependence of $\Delta\eta$ in Fig. 3c confirms that the $\tau^{\mathrm{fast}}$ component 
exclusively appears below T$_{N}$, originating therefore from the long-range magnetic order contribution to the anisotropy (black-red area). 
In sharp contrast, the $\tau^{\mathrm{slow}}$ component (white area), shown also in Fig. 3d at the fixed time delay of $\Delta\tau$=3.77ps, exhibits no change at low temperature, begins to drop at $\sim$T$_N$ and then fades out only at T$_s$=66K. 
$\tau^{\mathrm{slow}}$ (Fig. 3b) again shows a slowing down near T$_s$, thus originating from the electronic nematicity, since there is no long-range magnetic order at T$_N<$T$<$T$_S$,.

Fig. 3e compares the photoinduced $\Delta\eta$ dynamics in the x=0.047 sample versus temperature at two fixed time delays $\Delta\tau=$350fs and 3.77ps. While the $\Delta \eta$ amplitudes are different below T$_N$, they coincide above T$_N$, and then diminish 
at $T_s$. 
This salient feature allows us to determine the ratio between the contributions to the nematicity from the magnetic order versus other Ising-nematic contributions. 
Subtraction of the two amplitudes 
$\Delta \eta|_{\Delta\tau=350fs}$-$\Delta \eta|_{\Delta\tau=3.77ps}$ gives the {\em pure} magnetic order contribution, which sets in only below T$_N$ (blue shaded region).  
The sum of all other contributions is characterized by $\Delta \eta|_{\Delta\tau=3.77ps}$ (red shaded region), which extends up to T$_s$. The comparable magnitude between the two indicates a
substantial spin influence to the nematicity. 

Comparison of the fs-resolved nematicity and charge carrier dynamics reveals key couplings among various reservoirs. The transient temperature T$_{e}$ associated with the electron heat bath after ultrafast photoexcitation is proportional to the differential reflectivity profile, $\Delta$T$_{e}\propto \Delta$R/R($\Delta\tau$), for the first few ps \cite{25BeaurepairePRL95}. 
Below T$_N$, we introduce a reservoir temperature T$_m$ for the magnetic-nematic phase. In this regime both the nematic and magnetic order parameters contribute to the anisotropy, while the latter dominates the time evolution during the first few ps.
From the temperature dependence of the static $\eta$ (Fig. 1c) and the photoinduced changes $\Delta\eta$, we can extract the temporal profile of T$_m$ from the ultrafast data.  Fig. 4a shows T$_e$(t) (green squares) and T$_m$(t) (red circles) for the first 4 ps in the x=0.047 sample at an initial temperature 10K.
T$_e$(t) rises first during the ultrafast photoexcitation via laser heating of the electronic sub-system, followed by an increase of the magnetic-nematic reservoir temperature T$_m$ via heat transfer from the electron reservoir.
Most intriguingly, T$_m$ decays before equilibrating with T$_e$--before reaching T$_e$(t) profile--opposite to conventional magnetic metals \cite{25BeaurepairePRL95}. In nickel (inset, Fig, 4a), the magnetic and electron sub-systems first reach the same temperature and then lock together to decay towards an isothermal regime of the same heat bath temperature with phonons (T$_l$) (methods).  
This behavior in conventional metals is due largely to the relatively weak spin-phonon interaction compared to electron-phonon coupling, i.e., $g_{ml}/g_{el}\ll $1. 
However, the opposite trend in iron pnictides underpins an unusually strong spin-phonon coupling, i.e., $g_{ml}/g_{el}\gg  $1, as follows naturally from a magnetic-nematic state where the nematic order parameter mediates the coupling between spin and lattice degrees of freedom. 
For further quantitative understanding, 
we simulate the obtained ultrafast dynamics by a three-temperature (3T) model via a set of coupled differential equations of the energy flow for three reservoirs
after the pumping with rate P(t) by the 80fs Gaussian laser pulses over the excited volume (methods). 
The results of the calculation, shown in Fig. 4a, compare very well with the experiment, and reveal, particularly, a ``colossal" value of $g_{ml}/g_{el}$ $\sim$400, a ratio that is $\sim$10000 times larger than that in nickel ($\approx$0.04). 

A unified picture thus emerges for transient photo-driven cooperative processes in iron pnictides, as illustrated in Fig. 4b for low temperature T$< T_{s}, T_{N}$: ultrafast photoexcitations during 80 fs, shorter than the As-As optical phonon period, strongly alters the thermodynamic equilibrium between various reservoirs. 
I) During or
immediately following the pulse, electron-electron collisions lead to decoherence and
quick establishment of a quasi-thermal distribution of charge carriers after an electron thermalization time $\tau _{\mathrm{th}}^{e}$ $\sim$200fs. 
II) Next, the electrons cool down through energy transfer to the other baths. Spin-electron interaction gives rise to the softening of the spin-nematic order, which leads to the delayed rise $\Delta\tau_{rise}\sim$150fs of the thermalized magnetic-nematic order. 
III) The subsequent fast decay up to $\sim$2 ps of T$_m$ indicates lattice heating through strong spin-optical phonon coupling, which reaches an iso-thermal regime of lattice, electron, magnetic-nematic reservoirs $\sim$46K, still higher than that of the the equilibrium lattice temperature before the pump (10K).
IV) The slow recovery of the nematic order
proceeds within $\sim$20-100ps via the nematic fluctuation channel, 
and the system finally relaxes back to T=10K by heat diffusion between the laser pulse separation of 1 ms.     

\noindent$^{\dagger}$These authors contributed equally to this work.


\textbf{METHODS}\\
A brief summary is given here. More information on experimental techniques, data analysis and theoretical modeling for the static nematic order parameter $\phi$ and ultrafast dynamics is given in the supplementary section.

\textbf{Ba(Fe$_{1-x}$Co$_{x}$)$_2$As$_2$ growth and characterization.} 

Single crystals of $Ba(Fe_{1-x}Co_{x})_{2}As_{2}$, x=0.00 and 0.047, were grown out of a $(Fe_{1-y}Co_{y}){As}$ flux, using conventional high-temperature solution growth techniques.\cite{CanfieldARCMP10,
28NiPRB08} For the x=0.00, parent compound, small Ba chunks and FeAs powder were mixed together with 1:4 ratio. For the Co substituted compound, small Ba chunks, FeAs powder, and CoAs powder were mixed together according to the ratio Ba:FeAs:CoAs=1:3.75:0.25. Elemental analysis of the samples was performed using wavelength dispersive x-ray spectroscopy (WDS) in the electron probe microanalyzer of a JEOL JXA-8200 electron-microprobe to determine the real Co concentration.\cite{28NiPRB08} \\
\textbf{Laser-based static polarimetry measurements.} 
Here we develop a more sensitive laser probe from a Ti:Sapphire oscillator (1.55eV center wavelength, $\sim$100 fs pulse duration, and 80MHz repetition rate) that is frequency--doubled to 400nm (3.1eV), as shown in Fig. 1b. In the definition above and in the Jones Matrix notation, the incident $E$ field (s-polarized) and complex reflection from the sample are ${E_{in}}=(E_{s},E_{p})=(1,0)$ and ${E_{R}}=(1, \widetilde{\Theta} )$. The detection scheme of the ellipticity $\eta$ is achieved by passing the reflected beam through a quarter-wave plate (QWP), with its axis oriented at 45 degrees with respect to the vertical s-axis, and a polarization-dependent beam splitter to spatially separate the $s$ and $p$ polarized components. The anisotropy $\eta$ is calculated from 
\begin{equation}
2R\eta=\frac{R}{4}[\left | 
\left( \begin{array}{ccc}
1 & 0
\end{array} \right)
\left(\begin{array}{ccc}
1+i &1-i \\1-i 
 & 1+i
\end{array}\right)
\left(\begin{array}{ccc}
1\\ 
\widetilde{\Theta} 
\end{array}\right)
\right |^{2}-\left | \left(\begin{array}{ccc}
0&1 
\end{array}\right)\left(\begin{array}{ccc}
1+i &1-i \\1-i 
 & 1+i
\end{array}\right)\left(\begin{array}{ccc}
1\\ \widetilde{\Theta}
\end{array}\right)  
\right |^{2}],
\end{equation}
where $R$, defined as the reflection from the surface, is a measure of isotropic response equal to $I_s+I_p$. In the experiment, the probe beam is linearly polarized along the direction that produces maximum polarization activity. This is determined by the controlled experiment of probe polarization dependence which rotates the polarization plane long one of the tetragonal crystal vectors. The probe focus diameter is $\sim$200$\mu m$, which is smaller than the length of elongated strips from twinned domains \cite{7TanatarPRB10}.This allows optically probing the index anisotropy even in unstrained crystals using the highly-sensitive laser-based polarimetry, as clearly demonstrated in Fig. 1c.\\
\textbf{Ultrafast polarimetry measurements.} For time-resolved measurements, a Ti:Sapphire amplifier with  center 
wavelength of 800nm (1.55eV), pulse duration of $\sim$80 fs at the sample position, and 1kHz
repetition rate was separated into pump and probe beams. The probe was 
 frequency--doubled to 400nm (3.1eV) with a pulse duration of $\sim$120 fs. 
This photon energy has been show to measure the nematic order parameter $\phi$ from the static measurement shown in Fig. 2c.
This two--color pump--probe geometry was shown to minimize the 
contamination of  polarization signals during ultrafast time scales, e.g., by dichroic bleaching,  spectral weight transfer associated with the 
electronic phase transitions, etc (see, e.g., Ref. \cite{21StojchevskaPRB12} shows the near-infrared transient polarization singals persist far above all transition temperatures independent of doping).
Lattice and electron relaxation dynamics are revealed by pump-induced optical reflectivity change $\Delta R/R$. 
Small polarization or optical reflection changes were sampled as function of pump and probe delay by synchronously chopping the pump beam at 500 Hz and detecting the intensity change between consecutive pulses. \\
\textbf{Theoretical modeling of the nematic order parameter $\phi$.} 
Here the nematic order parameter $\phi$  is the scalar product of the two sublattice Neel vectors that make up the long range order below T$_N$: $\phi=<M_1 \times M_2>$. The important aspect of this Ising-spin nematic is that long range order of $\phi$ occurs already above the Neel temnperature, i.e. in a regime wher the magnetic order parameters still vanish $<M_1>=<M_2>$=0.\cite{15FernandesPRB12}
We follow the model extensively discussed in Ref. \cite{15FernandesPRB12} to obtain the static nematic order parameter $\phi$ as function of temperature (red line, Fig. 2b). In this approach, the existence of a doubly-degenerate magnetic ground state -- stripes with either $(\pi,0)$ or $(0,\pi)$ modulation -- combined with strong enough magnetic fluctuations gives rise to an Ising-nematic state which spontaneously breaks the tetragonal symmetry of the system already in the paramagnetic state (supplementary information).\\ 
\textbf{Theoretical modeling of ultrafast dynamics.} 
We model the obtained ultrafast dynamics with a three-temperature (3T) model. This is an extension of the two-temperature model that describes the electron (T$_e$) and lattice (T$_l$) thermal relaxation in metals by incorporating the T$_{m}$ for the magnetic-nematic phase.  The set of coupled differential equations of the energy flow are of the form below for three reservoirs with corresponding specific heat
coefficients C$_{e,m,l}$ and coupling constants g$_{ij}$ (=g$_{ji}$):
\begin{eqnarray}
C_{e} \frac{d}{dt} T_{e} = -g_{el} \left(T_{e}-T_{l}\right) -g_{em} \left(T_{e}-T_{m}\right)+P\left(t\right)\\
C_{l} \frac{d}{dt} T_{l} = -g_{el} \left(T_{l}-T_{e}\right) -g_{lm} \left(T_{l}-T_{m}\right)\\
C_{m} \frac{d}{dt} T_{m} = -g_{em} \left(T_{m}-T_{e}\right) -g_{lm} \left(T_{m}-T_{l}\right)
\end{eqnarray}
P(t) represents the pumping rate by the 80\textit{fs} Gaussian laser pulses over the excited volume. The electronic specific heat is given by: C$_{e}=\gamma_{e} T_{e}+C_{sc}$, where $\gamma = $ 6.1$\times10^{2}$ mJ/m$^{3}\cdot$K$^{2}$ and C$_{sc}=$ 2.475 $\times10^{2}$ mJ/m$^{3}\cdot$K is the albeit small, average superconducting contribution. The lattice specific heat, C$_{l}$, was determined from the Debye model with $\beta=10$ J/m$^{3}\cdot$K$^{4}$, while the magnetic-nematic specific heat C$_{m}$ was calculated from C$_{m}=C_{total} - C_{e} - C_{l}$. 
For the three temperature model, we numerically solved the system of equations simultaneously via fourth and fifth order Runge-Kutta formulas to make error estimates and adjust the 
time step accordingly, and extracted the coupling constants to be: g$_{el}=0.033\times10^{17}$ W/m$^{3}\cdot$K, g$_{em}=3.16\times 10^{17}$ W/m$^{3}\cdot$K, g$_{lm}=13.3\times10^{17}$ W/m$^{3}\cdot$K. Note that only the electron-lattice coupling have been reported before by time- and angle-resolved photoemission measurements and our extracted value is in agreement with the previous reports \cite{29RettigPRL12}. The inset shows the same simulation for nickel using the well-established parameters in the literature \cite{25BeaurepairePRL95}. The coupling constants for nickel is g$_{el}=8\times10^{17}$ W/m$^{3}\cdot$K, g$_{em}=6\times 10^{17}$ W/m$^{3}\cdot$K, g$_{lm}=0.3\times10^{17}$ W/m$^{3}\cdot$K.
Our results reveal, particularly, a ``colossal" value of $g_{ml}/g_{el}$ $\sim$400, a ratio that is $\sim$10000 times larger than that in Ni.     


\begin{thebibliography}{1}

\bibitem{7TanatarPRB10} Tanatar, M. A. \textit{et.al.}, 
{\em Uniaxial-strain mechanical detwinning of $CaFe_{2}As_{2} and BaFe_{2}As_{2}$ crystals: Optical and transport study},
\textit{Phys. Rev. B}  \textbf{81}, 184508 (2010).


\bibitem{1ChuScience10} Chu, J. \textit{et.al.}, 
{\em In-Plane Resistivity Anisotropy in an Underdoped Iron Arsenide Superconductor},
\textit{Science} \textbf{329}, 824 (2010).


\bibitem{3YingPRL11} Ying, J. J. \textit{et.al.}, 
{\em Measurements of the Anisotropic In-Plane Resistivity of Underdoped FeAs-Based Pnictide Superconductors},
\textit{Phys. Rev. Lett.}  \textbf{107}, 067001 (2011).

\bibitem{4YiPNAS11} Yi, M. \textit{et.al.}, 
{\em Symmetry-breaking orbital anisotropy observed for detwinned $Ba(Fe_{1-x}Co_{x})_{2}As_{2}$ above the spin density wave transition},
\textit{Proc. Natl. Acad. Sci.}  \textbf{108}, 17 (2011).

\bibitem{5NakajimaPNAS11} Nakajima, M. \textit{et.al.}, 
{\em Unprecedented anisotropic metallic state in undoped iron arsenide $BaFe_{2}As_{2}$ revealed by optical spectroscopy},
\textit{Proc. Natl. Acad. Sci.}  \textbf{108}, 30 (2011).

\bibitem{6DuszaEPL11} Dusza, A. \textit{et.al.}, 
{\em Anisotropic charge dynamics in detwinned $Ba(Fe_{1-x}Co_{x})_{2}As_{2}$},
\textit{Eur. Phys. Lett.}  \textbf{93}, 37002 (2011).

\bibitem{chu2012} Chu, J., Kuo, H., Analytis, J. G., and Fisher, I.R., 
{\em Divergent nematic susceptibility in an iron arsenide superconductor},\textit{Science}, 
\textbf{337}, 710 (2012).

\bibitem{Gegenwart13} Jiang, S., Jeevan, H. S., Dong, J. \&  Gegenwart, P. 
{\em Thermopower as sensitive probe of electronic nematicity in iron pnictides},
\textit{Phys. Rev. Lett.}  \textbf{110}, 067001 (2013).


\bibitem{16KasaharaNature12} Kasahara, S. \textit{et.al.}, 
{\em Electronic nematicity above the structural and superconducting transition in $BaFe_{2}(As_{1-x}P_{x})_{2}$},
\textit{Nature}  \textbf{486}, 382 (2012).

\bibitem{9SiPRL08} Si, Q. \& Abrahams, E.
{\em Strong Correlations and Magnetic Frustration in the High $T_{c}$ Iron Pnictides},
\textit{Phys. Rev. Lett.} {\bf 101}, 076401 (2008). 


\bibitem{10FangPRB08} Fang, C., Yao, H., Tsai, W., Hu, J. \& Kivelson, A.
{\em Theory of electron nematic order in LaFeAsO},
\textit{Phys. Rev. B} {\bf 77}, 224509 (2008). 

\bibitem{11XuPRB08} Xu, C., Muller, M. \& Sachdev, S.
{\em Ising and spin orders in the iron-based superconductors},
\textit{Phys. Rev. B} {\bf 78}, 020501 (2008). 

\bibitem{DevereauxPRB10} C.-C. Chen, J. Maciejko, A. P. Sorini, B. Moritz, R. R. P. Singh, \& T. P. Devereaux
{\em Orbital order and spontaneous orthorhombicity in iron pnictides},
\textit{Phys. Rev. B} {\bf 82}, 100504 (2010).


\bibitem{13FernandesPRL10} Fernandes, R. M. \textit{et.al.}, 
{\em Effects of Nematic Fluctuations on the Elastic Properties of Iron Arsenide Superconductors},
\textit{Phys. Rev. Lett.}  \textbf{105}, 157003 (2010).


\bibitem{15FernandesPRB12} Fernandes, R. M., Chubukov, A. V., Knolle, J., Eremin, I. \& Schmalian, J.
{\em Preemptive nematic order, pseudogap, and orbital order in the iron pnictides},
\textit{Phys. Rev. B} {\bf 85}, 024534 (2012). 

\bibitem{FradkinARCMP10} Fradkin, E., Kivelson S. A., Lawler, M. J., Eisenstein, J. P.,\& Mackenzie, A. P.
{\em Nematic Fermi Fluids in Condensed Matter Physics},
\textit{Annu. Rev. Cond. Mat. Phys.} {\bf 1}, 153-178 (2010).

\bibitem{KimPRB11} Kim, M. G.  \textit{et.al.}, 
{\em Character of the structural and magnetic phase transitions in the parent and electron doped BaFe$_{2}$As$_{2}$ compounds},
\textit{Phys. Rev. B} {\bf 83}, 134522 (2011). 

\bibitem{Merteljprl2009} Mertelj, T.,  \textit{et.al.}, 
{\em Distinct Pseudogap and Quasiparticle Relaxation Dynamics in the Superconducting State of Nearly Optimally Doped $SmFeAsO_{0.8}F_{2}$ Single Crystals}, \textit{Phys. Rev. Lett}  \textbf{102}, 117002 (2009). 

\bibitem{18MansartPRB10} Mansart, B. \textit{et.al.}, 
{\em Ultrafast transient response and electron-phonon coupling in the iron-pnictide superconductor $Ba(Fe_{1-x}Co_{x})_{2}As_{2}$},
\textit{Phys. Rev. B}  \textbf{82}, 024513 (2010).

\bibitem{19ChiaPRL10} Chia, E. E. M. \textit{et.al.}, 
{\em Ultrafast Pump-Probe Study of Phase Separation and Competing Orders in the Underdoped $(Ba,K)Fe_{2}As_{2}$ Superconductor},
\textit{Phys. Rev. Lett.}  \textbf{104}, 027003 (2010).

\bibitem{20TorchinskyPRL10} Torchinsky, D. H., Chen, G. F., Luo, J. L., Wang, N. L. \& Gedic, N. {\em Band-dependent Quasiparticle Dynamics in Single Crystals of the $Ba_{0.6}K_{0.4}Fe_{2}As_{2}$ Superconductor Revealed by Pump-Probe Spectroscopy},
\textit{Phys. Rev. Lett.} {\bf 105}, 027005 (2010). 

\bibitem{21StojchevskaPRB12} Stojchevska, L., Mertelj, T., Chu, J., Fisher, I. R. \& Mihailovic, D. {\em Doping dependence of femtosecond quasiparticle relaxation dynamics in $Ba_{0.6}K_{0.4}Fe_{2}As_{2}$ single crystals: Evidence for normal-state nematic fluctuations}, \textit{Phys. Rev. B} {\bf 86}, 024519 (2012).

\bibitem{Kim2012} Kim, K. W. \textit{et.al.}
{\em Ultrafast transient generation of 
spin-density-wave order in the normal state of 
BaFe$_2$As$_2$ driven by coherent lattice vibrations}, 
\textit{Nature Mater.} \textbf{11}, 497 (2012).

\bibitem{LiNature13} Li, T. \textit{et.al.},
{\em Femtosecond switching of magnetism via strongly correlated spin¨Ccharge quantum excitations},
\textit{Nature} {\bf in press}, DOI 10.1038/nature11934 (2013).

\bibitem{LiPRL12} Wang, J.\textit{et.al.},
{\em Ultrafast Softening in InMnAs},
\textit{Physica E} {\bf 20}, 412 (2005). 

\bibitem{CanfieldARCMP10} Canfield, P. C., \& Bud'ko, S. L.
{\em FeAs-Based Superconductivity: A Case Study of the Effects of Transition Metal Doping on $BaFe_{2}As_{2}$},
\textit{Annu. Rev. Cond. Mat. Phys.} {\bf 1}, 27-50 (2010). 


\bibitem{25BeaurepairePRL95} Beaurepaire, E., Merle, J. C., Daunois, A. \& Bigot, J. Y. {\em Ultrafast Spin Dynamics in Ferromagnetic Nickel},
\textit{Phys. Rev. Lett.} {\bf 76}, 4250 (1995).

\bibitem{23NandiPRL10} Nandi, S. \textit{et.al.}, 
{\em Anomalous Suppression of the Orthorhombic Lattice Distortion in Superconducting $Ba(Fe-{1-x}Co_{x})_{2}As_{2}$ Single Crystals},
\textit{Phys. Rev. Lett.}  \textbf{104}, 057006 (2010).






\bibitem{28NiPRB08} Ni, N. \textit{et.al.}, 
{\em Effects of Co substitution on thermodynamic and transport properties and anisotropic
$H_{c2}$ in $Ba(Fe_{1-x}Co_{x})_{2}As_{2}$ single crystals}, \textit{Phys. Rev. B}  \textbf{78}, 214515 (2008).


\bibitem{29RettigPRL12} Rettig, L. \textit{et.al.}, 
{\em Ultrafast Momentum-Dependent Response of Electrons in Antiferromagnetic
EuFe$_2$As$_2$ Driven by Optical Excitation}, \textit{Phys. Rev. Lett}  \textbf{108}, 097002 (2012).










\end{thebibliography}


\begin{addendum}
 \item This work was supported in part by the Ames Laboratory's LDRD program  (sample characterization and computational studies), by the U.S. Department of Energy, Office of Science, Basic Energy Sciences, Materials Science and Engineering Division (materials synthesis), by the National Science Foundation under award DMR-1055352 (laser spectroscopy).  Ames Laboratory is operated for the U.S. DOE by Iowa State University under contract DE-AC02-07CH11358. I.E.P was supported by the European Union's Seventh Framework Programme (FP7-REGPOT-2012-2013-1) under grant agreement No. 316165 and by the EU Social Fund and National resources through the THALES program NANOPHOS. 

\item[Author Contributions]
T.L., A.P., and J.W. performed the experimental measurements and collected the data. 
R.M.F. and J.S. performed theoretical modeling of the nematic order parameter $\phi$.
A.P., T.L., J.W. analyzed and performed theoretical modeling of ultrafast dynamics with help from I.E.P. 
Single crystal synthesis, basic characterization and analysis were done by S.R., P.C.C. and S.L.B.
All authors discussed results together. 
J.W. designed the experiment and wrote the paper, with help from all authors. 

 \item[Competing Interests] The authors declare that they have no
competing financial interests.
 \item[Correspondence] Correspondence and requests for materials
should be addressed to J.W.~(email: jgwang@iastate.edu).
\end{addendum}

\begin{figure}
\includegraphics[scale=0.5]{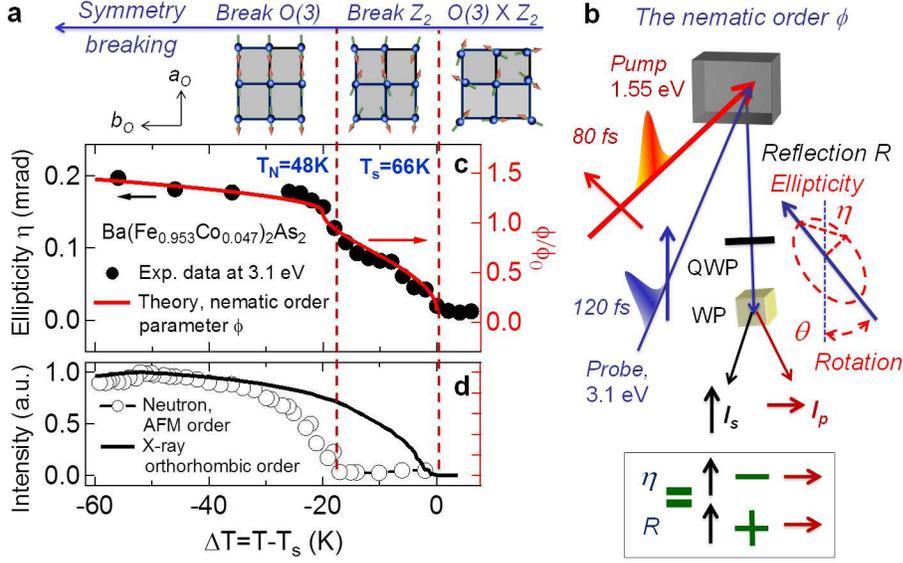}
\caption{\scriptsize{A laser-based polarimetry technique measures the ellipticity $\eta$, which is directly proportional to the Ising-nematic order parameter $\phi$, in $\mathrm{Ba(Fe_{1-x}Co_{x})_{2}As_{2}}$ systems. \textbf{a}, Schematics of the establishment of the Ising-nematic phase via spontaneous tetragonal symmetry breaking, illustrated in the ab plane. At high temperatures, $O(3)$ spin-rotational symmetry is unbroken, as well as the Ising-nematic $Z_2$ symmetry associated with the doubly-degeneracy of the magnetic ground state, corresponding to magnetic stripes with either (0, $\pi$) or ($\pi$, 0) modulation. Below the tetragonal-to-orthorhombic transition temperature $T_s$, this emergent $Z_2$ symmetry is broken, but the $O(3)$ symmetry is preserved until $T_N \leq T_s$, when long-range magnetic order sets in. At $T_s$ the system also develops orbital and structural order.
\textbf{b}, As illustrated, the ellipticity of the reflected beam is detected by passing the reflected beam through a quarter-wave plate (QWP), with its axis oriented at 45 degrees with respect to the vertical s-axis, and a polarization-dependent beam splitter (wollaston prism, WP) to spatially separate the $s$ and $p$ polarized components (methods). The anisotropy component $\eta$ is measured from $I_s-I_p$ and the isotropic contribution $R$ from $I_s+I_p$. \textbf{c}, The measured temperature-dependent $\eta$ (black solid dots) and the calculated normalized nematic order parameter $\phi/\phi_{0}$ for the x=0.047 sample agree very well. The coupling between the nematic and magnetic order parameters -- black empty dots shown in \textbf{d}, extracted via neutron scattering -- is manifested by the hump at $T_N=48$K. The coupling between $\phi$ and the lattice gives rise to a non-zero orthorhombicity at $T_s=66$K, as found by x-ray scattering -- black line in \textbf{d}.}}
\end{figure}

\begin{figure}
\includegraphics[scale=0.7]{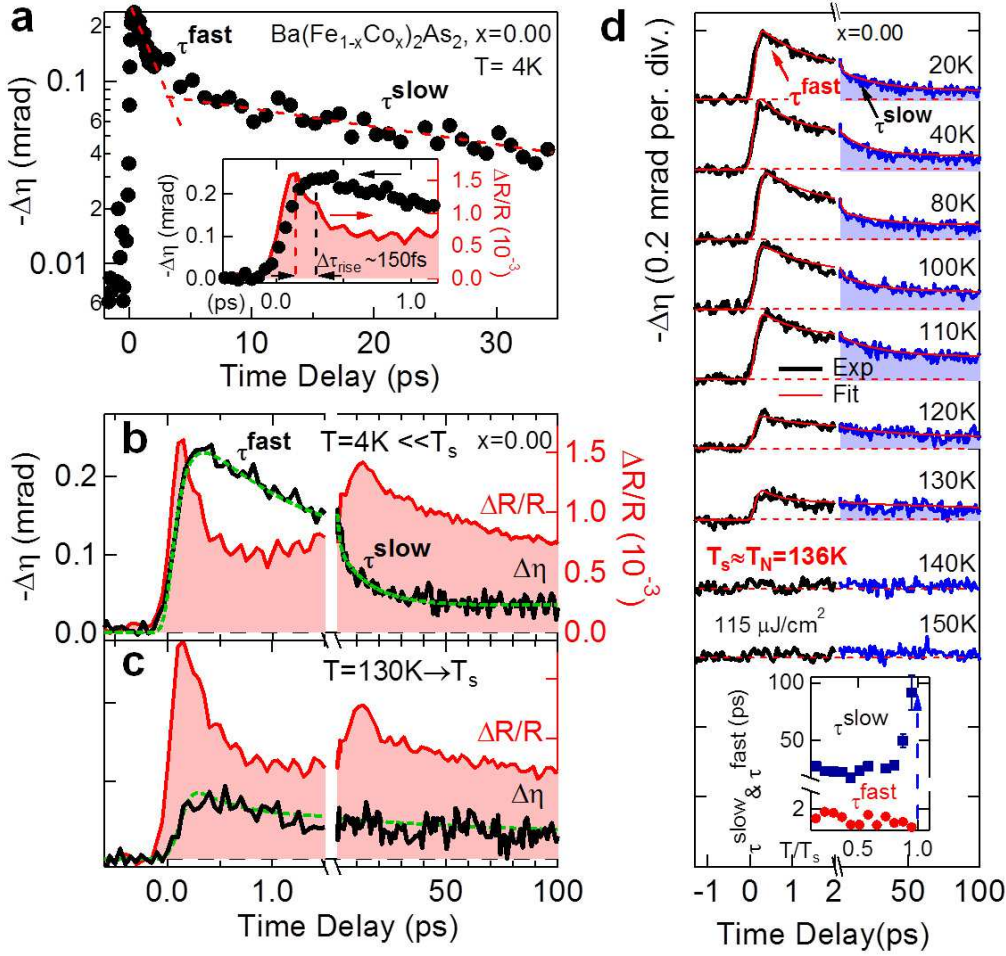}
\caption{\scriptsize{Ultrafast photoinduced dynamics of the nematic order parameter in the parent compound BaFe$_2$As$_2$. 
\textbf{a}, Ultrafast photoinduced change in the ellipticity $\Delta\eta$ at $T=4$K for 35 ps and for the first 1.2 ps of time delay (black dots, inset). 
$\Delta\eta$ exhibits a double-exponential time-dependent recovery with time constants $\tau^{\mathrm{fast}}=$1.2ps and $\tau^{\mathrm{slow}}=28$ps  (red dash lines). The inset reveals a delayed rise $\Delta\tau_{rise}$ of $\Delta\eta$ (black dots) compared to $\Delta R/R$ (red shade), which indicates a faster electron thermalization $\sim150$fs before the softening of the nematic order.
\textbf{b} and \textbf{c}, the photoinduced ellipticity $\Delta\eta$ (back lines) and $\Delta R/R$ (red shades) for the first 1.5 ps and for the extended time scales of 100 ps (split axis) at two temperatures 4K (\textbf{b}) and 130K (\textbf{c}). 
Note that the $\Delta R/R$ data is superimposed to periodical oscillations of 120 fs caused by an As-As optical phonon \cite{18MansartPRB10}.
\textbf{d}, Detailed temperature dependence of the photoinduced ellipticity $\Delta\eta$ for the first 2 ps (black lines) and extended 100 ps (blue, split axis), shown together with double-exponential fittings (red lines). The extracted $\tau^{\mathrm{fast}}$ (red circles) and $\tau^{\mathrm{slow}}$ (blue diamonds) are presented in the inset as a function of the normalized temperature $T/T_{s}$. Note the critical slowing down in the $\tau^{\mathrm{slow}}$ relaxation time at $T_s$ (blue dashed line).}}
\end{figure}

\begin{figure}
\includegraphics[scale=0.8]{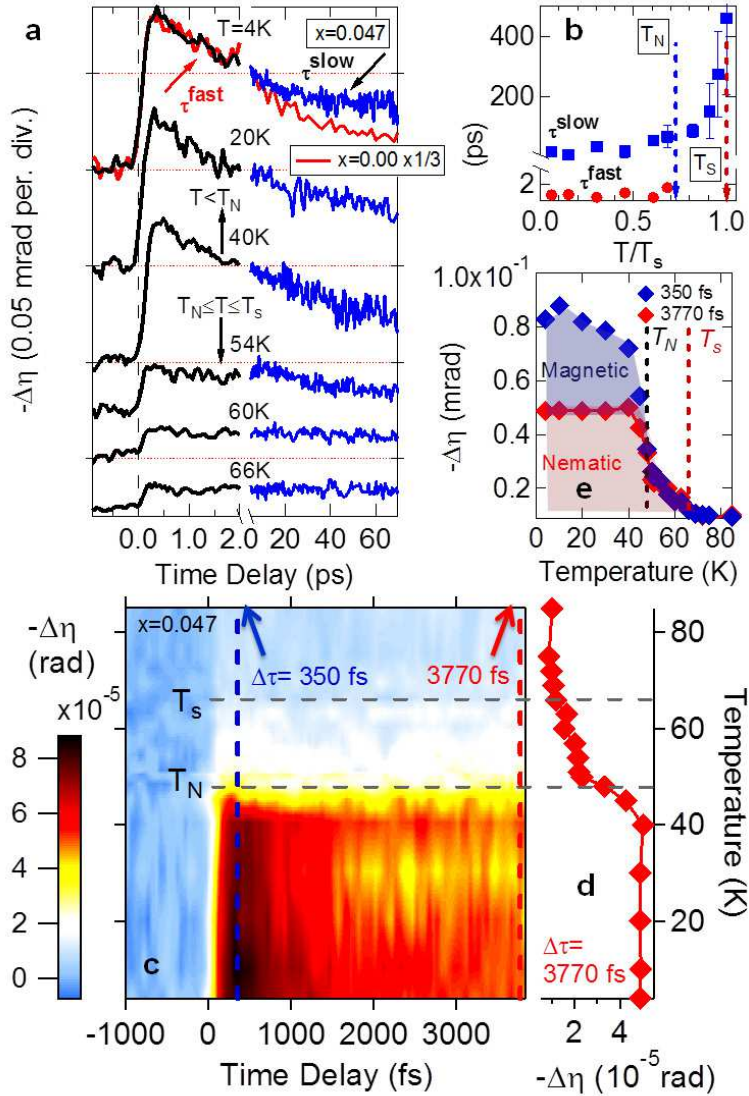}
\caption{\scriptsize{\textbf{Ultrafast photoinduced dynamics of the nematic order parameter in (under-doped) Ba(Fe$_{0.953}$Co$_{0.047}$)$_2$As$_2$}.
\textbf{a} Detailed temperature dependence of the ultrafast photoinduced ellipticity change $\Delta\eta$ for x=0.047  samples for the first 2 ps  (black lines) and for extended time scales of 100 ps (blue lines, split axis), shown together with the 4K ellipticity trace for x=0 sample (red lines). The pump fluence used for all six traces is 115$\mu$J/cm$^{2}$. 
\textbf{b} The extracted $\tau^{\mathrm{fast}}$ (red circles) and $\tau^{\mathrm{slow}}$ (blue squares) with error bars, defined as the s.d. in the fitting, are presented as a function of the normalized temperature $T/T_{S}$. 
Note the critical slowing down in the $\tau^{\mathrm{slow}}$ relaxation time at $T_S$ (red dashed line).
\textbf{c} A false color image of the time-resolved ellipticity change $\Delta\eta$ for the x=0.047 sample as function of temperature, showing distinct transitions at both $T_N$ and $T_S$. 
 \textbf{d} Temperature dependence of the photoinduced $\Delta\eta$ at $\Delta\tau$=3770fs. \textbf{e} Temperature dependence of the photoinduced $\Delta\eta$ at two time intervals $\Delta\tau$=350fs (blue solid diamonds) and 3770fs (red solid diamonds), following photoexcitation. The blue and red  shaded regions show the inferred contributions to the nematicity coming from the magnetic order parameters (marked as magnetic) and other Ising-nematic contributions (marked as nematic), respectively.}} 
\end{figure}

\begin{figure}
\includegraphics[scale=0.6]{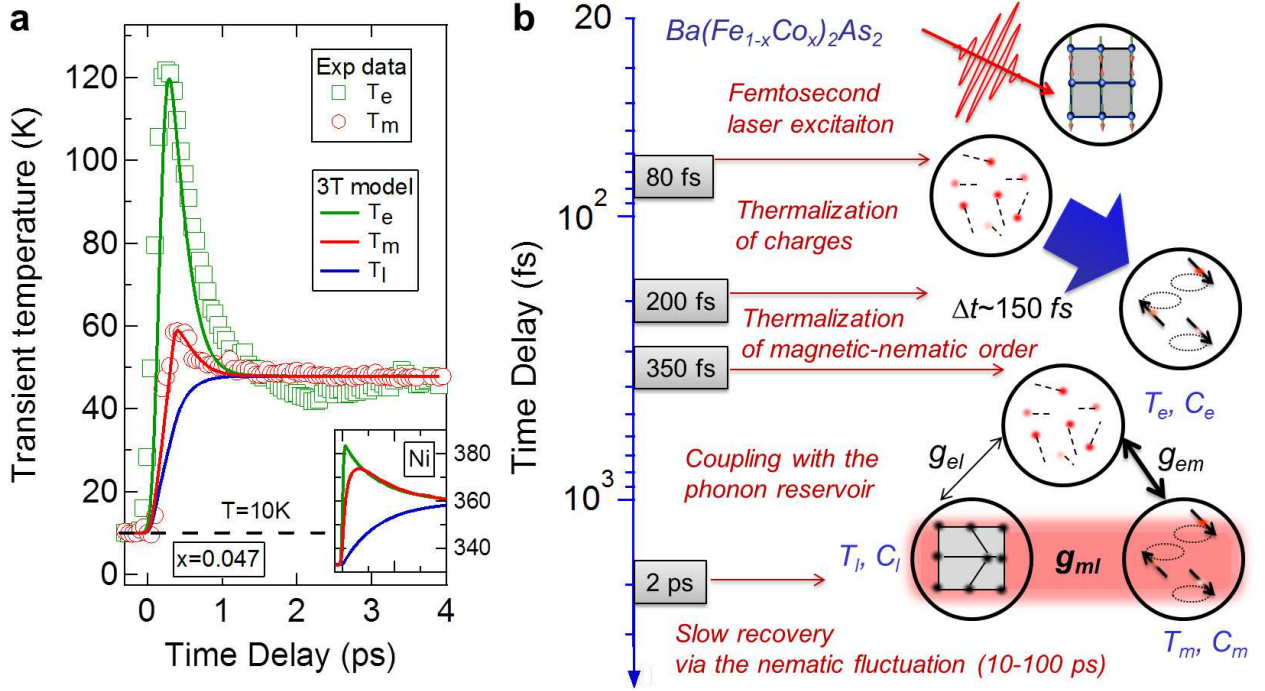}
\caption{\scriptsize{Ultrafast photoinduced transient cooperative processes for the magnetic-nematic, electronic, and lattice reservoirs in iron pnictides. 
\textbf{a}, the measured and calculated (3T model) ultrafast transient temperature changes for the electron ($T_e$), lattice ($T_l$) and magnetic-nematic ($T_{m}$) heat baths at $T=10$K. Shown together in the inset is the result from elemental nickel for comparison ($T=300$K, $T_c=631$K). \textbf{b}, A unified picture emerges for transient photo-driven cooperative processes in iron pnictide systems with the respective time scales extracted from our analysis. The illustration refers to low temperatures $T<T_N , T_s$ under the 80 fs photoexcitation. It depicts, in this order: the electron thermalization and cooling, the magnetic-nematic order thermalization, the lattice heating through strong spin-optical phonon coupling, the slow recovery of the nematic order via its own fluctuation diverging at T$_s$, and the final relaxation back to $T=10$K by heat diffusion. }}
\end{figure}

\end{document}